\documentclass[aps,pre,reprint,amsmath,amssymb]{revtex4-2}
\usepackage{graphicx,subfigure}
\usepackage{dcolumn}
\usepackage{epstopdf}
\usepackage{color}
\usepackage{multirow}
\usepackage{upgreek}
\usepackage[hidelinks]{hyperref}

\usepackage{bm}

\makeatletter
\def\btt#1{\texttt{\@backslashchar#1}}%
\DeclareRobustCommand\bblash{\btt{\@backslashchar}}%
\makeatother

\begin{document}

\title{Nontrivial electrophoresis of silica micro and nanorods in a nematic liquid crystal}

\date{\today}
\author{Muhammed Rasi M$^{1}$, Archana S$^{1}$, Ravi Kumar Pujala$^{2}$,  and Surajit Dhara$^{1}$}
\email{Email: surajit@uohyd.ac.in} 
\affiliation{$^1$School of Physics, University of Hyderabad, Hyderabad-500046, India\\
$^{2}$Department of Physics, Indian Institute of Science Education and Research, Tirupati, Andhra Pradesh 517507, India
}

\begin{abstract}

We study  DC and AC electrophoresis of silica micro and nanorods in a thin film of a nematic liquid crystal.  These particles induce virtual topological defects and also demonstrate nontrivial electrophoresis in a nematic liquid crystal. We  measure several nonlinear electrophoretic mobility coefficients and compare with those calculated theoretically. We demonstrate a competing effect of the elastic and electrostatic torques that arises due to tilting of the rods with respect to the liquid crystal director. A basic theory describing this effect allows us to measure the effective polarisability of the rods. Our approach is simple and applicable to a wide variety of asymmetric and polarisable particles. 
% We report experimental studies on DC and AC electrophoresis of silica micro and nanorods in a  thin film of a nematic liquid crystal with negative dielectric anisotropy. The DC electrophoresis is studied in cells with in-plane stripe electrodes and the AC electrophoresis is studied in Hele-Shaw cells  with homogeneous alignment of the director. In both cells the direction of electric field was perpendicular to the director. In cells with in-plane electrodes the rods propel at an angle with respect to the director. We measure nonlinear electrophoretic mobility coefficients and compare with those calculated theoretically. In Hele-Shaw cells, the rods propel parallel to the director and also tilt along the field direction which is explained by a simple theory based on the competing effect of the elastic and electrostatic torques. It allows us to estimate the effective polarisability of the rods.
 %Our experiments demonstrate a simple method for measuring electrophoretic mobilities as well as effective polarisability of the rods which is applicable to a wide variety of asymmetric  and polarisable particles.  
 
\end{abstract}

\preprint{HEP/123-qed}
\maketitle

\section{Introduction}
Study on the self-propulsion and active transport of microscopic objects in fluids is not only fundamentally important but also essential for technological development in diverse areas like targeted delivery, microrobotics, controlled assembly, transport and reactions \textcolor{blue}{\cite{drop,alex,div2,rc}}.
 Particles in a fluid can be transported by transducing  the energy of external electric, magnetic fields, light or by chemical reaction. Collective motion of such particles provides many useful insights of problems in non-equilibrium physics and active matter~\textcolor{blue}{\cite{cb,sje,rg1,rg2}}. Electrophoresis is one of the commonly used methods for transporting microparticles in isotropic fluids such as in water~\textcolor{blue}{\cite{morgan,ramos}}. The electrophoretic velocity is usually 
 linear or nonlinear with the applied electric field. In linear electrophoresis, the applied electric field is low and  the surface of the dielectric particles accumulate charge screend by counter ions, forming an electric double layer. The external electric field ($E$) triggers the motion of the charges giving rise to a slip velocity and consequently the particles move with a velocity given by Helmholtz-Smoluchowski (HS) equation $v=\epsilon_0\epsilon\zeta E/\eta $, where $\eta $ is the dynamic fluid viscosity and $\zeta$ is the surface potential~\textcolor{blue}{\cite{na}}. For particles with low Dukhin number, $Du=\sigma_P/(\sigma_F r) <<1$ ($\sigma_P$: particle surface conductivity and $\sigma_F$: fluid conductivity and $r$: radius) the velocity is independent of the size and shape of the particles as predicted by the HS equation~\textcolor{blue}{\cite{na}}. However if the electric field strength is increased the electric potential at the particle's surface is increased by a quantity $\propto E^2$ due to field induced polarisation of electric double layers; a phenomena known as Stotz-Wein effect~\textcolor{blue}{\cite{na,s,t}}. In this case $Du>>1$, and the  velocity becomes nonlinear is given by $v=\frac{\epsilon_0\epsilon}{\eta} (\zeta E+\beta(r)E^n)$, where the index depends on the Peclet number $Pe=rv/D$, $D$ being the diffusion coefficient of the ions. Shape asymmetric particles with polarisable surface also exhibit induced charge electrophoresis in which the applied field acts on the ionic charge it induces around the surface and the velocity of the particle is proportional to $ E^2$~\textcolor{blue}{\cite{tm,baz1,baz2,baz3}}.

\begin{figure}[!ht]
\center\includegraphics[scale=0.3]{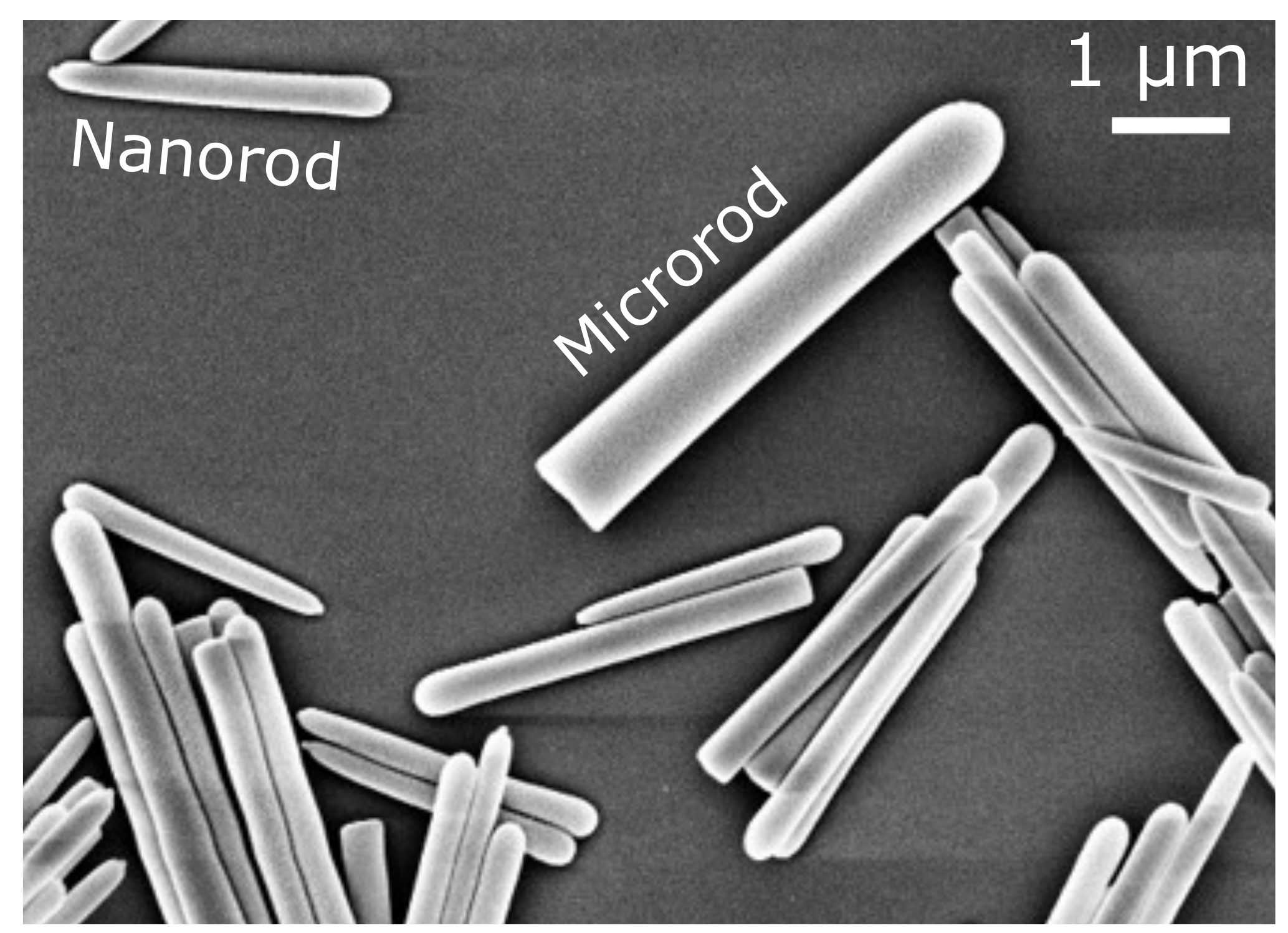}
\caption{Scanning electron microscope (SEM) image of mixture of silica nanorods  and  microrods.
 \label{fig:fig1} }
\end{figure}

  When the fluid is anisotropic, like liquid crystals, the electrophoresis is markedly different because of the anisotropic electrical properties of the medium and the induced topological defects by the particles~\textcolor{blue}{\cite{im1,is1}}. The electrophoresis of spherical dielectric particles without any charge is nonlinear as the electric field creates charge separation across the particles and drives the surrounding fluid flow by the electrostatic force of the same field~\textcolor{blue}{\cite{oleg1}}. The fore-aft symmetry of the electroosmotic flow is broken for particles with dipolar director symmetry (nucleating point defects) and the particles propel along the director with a velocity, $v\propto E^2$. This is known as liquid crystal-enabled electrophoresis (LCEEP)~\textcolor{blue}{\cite{oleg,od,od1}}. The fore-aft symmetry is preserved for quadrupolar particles ( nucleating boojums and Saturn rings) and the particles do not propel~\textcolor{blue}{\cite{oleg2}}. In general the velocity-field relationship of a particle in nematic LCs is tensorial and can be expressed as~\textcolor{blue}{\cite{oleg}}
  
  \begin{equation}
  v_{i}=\mu_{1ij}E_{j}+\beta_{ijk}E_{j}E_{k}+\mu_{3ijkl} E_{j}E_{k}E_{l}
  \end{equation}
where $\mu_{1ij}$, $\mu_{3ijkl}$ and $\beta_{ijk}$ are the tensorial mobility coefficients. The first and the third terms demonstrate the classical electrophoresis of charged particles and the second term is attributed to the liquid crystal-enabled electrophoresis (LCEEP).

  While the electrophoresis of spherical particles in NLCs has been studied, recent advances in colloid science has made it possible to make microparticles with nontrivial shapes and electrophoresis of such particles is largely unexplored. Here, we report experimental studies on the electrophoresis of two  silica rods with different aspect ratio in a NLC. We experimentally measure linear and nonlinear electrophoretic mobilities and corroborate the results with theory. We show that in Hele-Shaw cells the rods are tilted due to a competing effect of elastic and electrostatic torques. The experimental results are corroborated with a simple theory which allows us to estimate effective polarisability oof the silica rods.

%%%%%%%%%%%
\section{Experiment}
Silica rods were synthesized following a wet chemical method~\textcolor{blue}{~\cite{Kuijk}}. Initially, 3g of polyvinylpyrrolidone (PVP) was dissolved in 30 ml of 1-pentanol by continuous shaking and sonication. After PVP had been completely dissolved, 3 ml of ethanol (100\%, Interchema), 0.84 ml Milli Q water, and 0.2 ml aqueous sodium citrate dyhydrate (0.17M, 99\%, Sigma-Aldrich) were added sequentially. The reaction mixture was thoroughly mixed to avoid any bubble formation. Finally, 0.3 ml of TEOS (98\%, Sigma-Aldrich) was added to the reaction mixture, gently shaken, and the reaction was allowed to continue for a day. Finally, the product mixture was centrifuged, cleaned many times and fractionated to achieve rods with desired aspect ratio. Here, we considered rods with two different aspect ratios (see Table-\ref{table:1}). The particles with average length and diameter, $L=3~\upmu$m and $D= 200$ nm, are designated as nanorods and particles with $L=5 ~\upmu$m and $D=700$ nm are designated as microrods for in the discussion (see Table-\ref{table:1}). The scanning electron microscope (SEM) image of a nano- and a microrod is shown in Fig. \ref{fig:fig1}. The average Zeta potentials ($\zeta$) of the nano- and microrods in ethanol measured  by a particle size analyser (Litesizer 500, Anton Paar) are -13 mV and -15 mV, respectively (see Table-\ref{table:1}).

Further the silica rods were coated with octadecyldimethyl-3-trimethoxysilyl propyl-ammonium chloride (DMOAP) for inducing homeotropic anchoring of the liquid crystal molecules. We prepared a dilute suspension ($<$0.1 wt\%) of the rods into a negative dielectric anisotropy nematic liquid crystal MLC-6608 (Merck).
Cells with in-plane stripe electrodes were prepared by two parallel copper tapes as shown in Fig.\ref{fig:fig4}(a). Hele-Shaw cells were prepared by 
two indium-tin-oxide (ITO) coated glass plates as shown in Fig.\ref{fig:fig5}(a). Both the cells were treated for homogeneous alignment of the director. We used a DC voltage source (APLAB) for DC electrophoresis and a function generator (Tektronix-AFG 3102) connected to a voltage amplifier for AC electrophoresis. An infrared (1064 nm) laser tweezers setup mounted on optical polarizing microscope (Nikon Eclipse Ti-U) was used for particle manipulation \textcolor{blue}{\cite{Rasi,Zuhail,Zuhail1}}. The position of the moving rods was tracked from recorded movies by using a suitable particle tracking program. 

\begin{table}[h!]
\centering
\begin{tabular}{ |p{1.5cm}|p{1.4 cm}|p{1.6 cm}|p{1.6cm}|p{1.6cm}| }
 \hline
Particles & Average Length, $L$ ($\upmu$m) &  Average Diameter, $D$ (nm)& Aspect ratio ($L/D$) & Zeta potential $\zeta$ (mV) \\
 
\hline
 Nanorods       & 3 & 200 & 15 &-13$\pm 1$    \\
\hline
 Microrods     &5 & 700 & 7 &-15$\pm 1$   \\
\hline
\end{tabular}

\caption{ Average length ($L$) and diameter ($D$), aspect ratio and Zeta potential of the rods.}
\label{table:1}
\end{table}

\begin{figure}[!ht]
\center\includegraphics[scale=0.3]{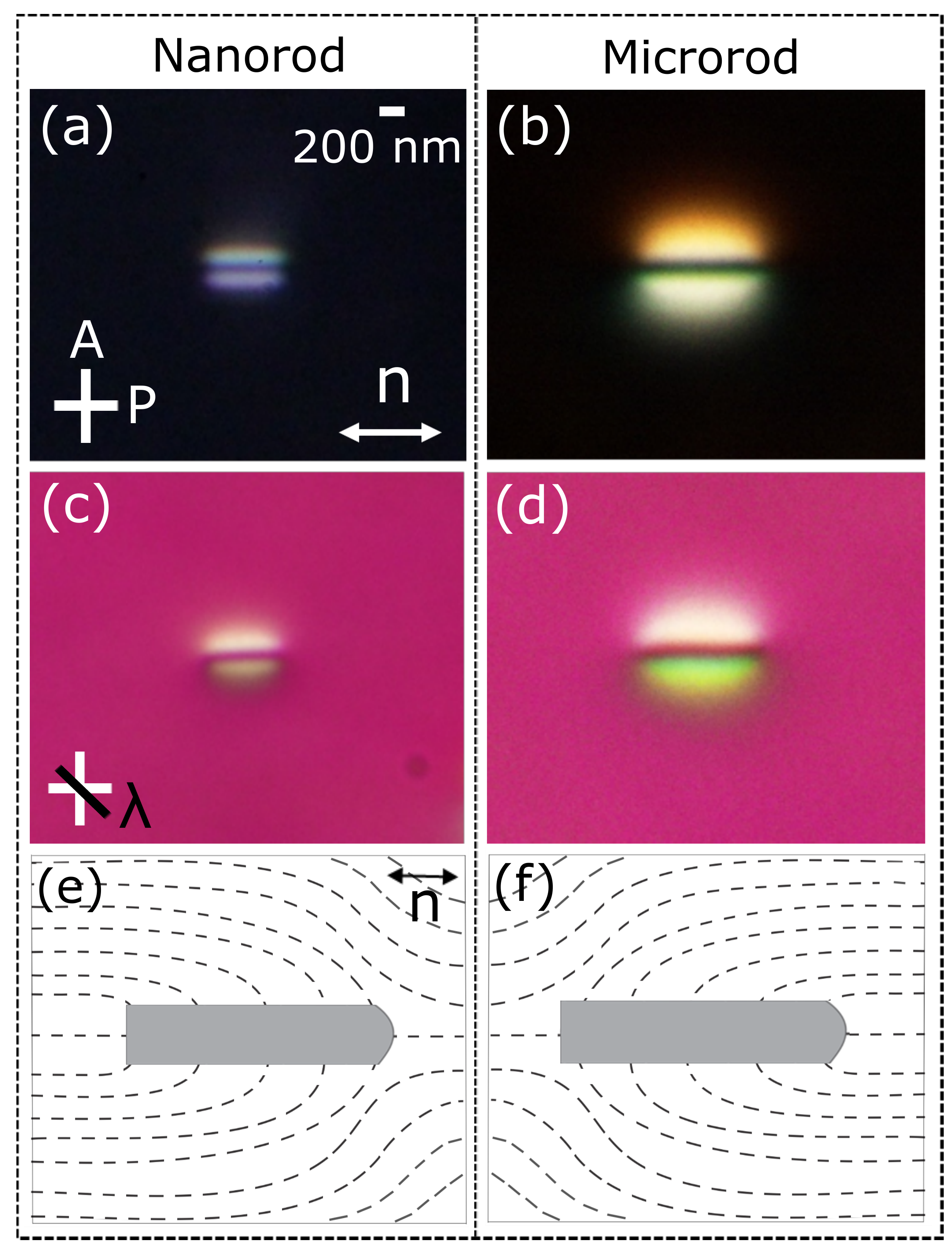}
\caption{ Rods aligned parallel to the director: Polarising optical photomicrographs (POM) of a DMOAP coated (a) nanorod and (b) microrod in MLC-6608 NLC. Double-headed arrow at the right bottom indicates director ${\bf {\hat{n}}}$. (c,d) Photomicrographs with a $\lambda$-plate (slow-axis) oriented at 45$^{\circ}$. (e,f) Schematic diagram of two possible director configurations around a micro or nanorod. }
 \label{fig:fig2} 
\end{figure}

%%%%%%%%%%%%%%%
\section{Results and discussion}

\begin{figure}[!ht]
\center\includegraphics[scale=0.25]{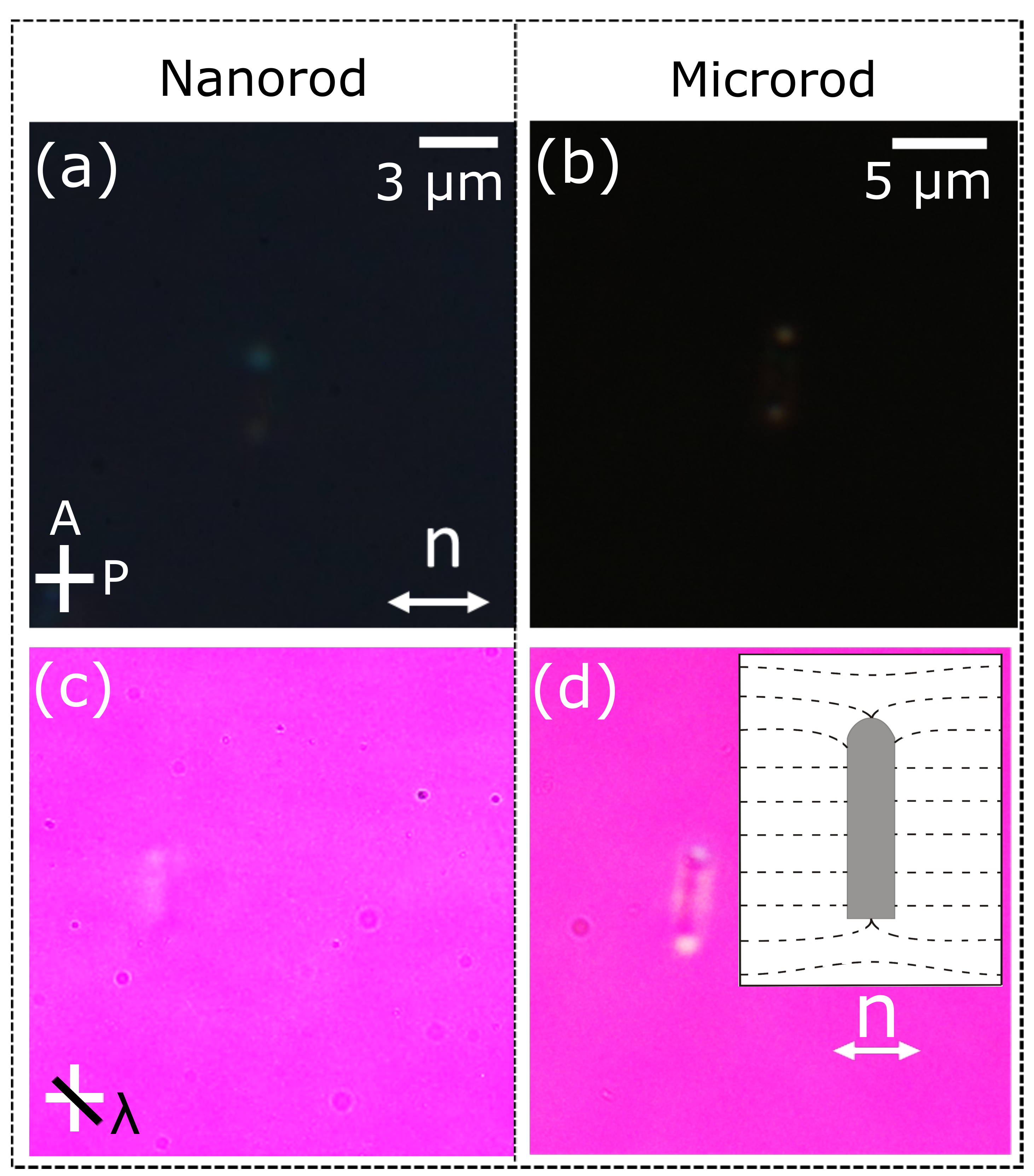}
\caption{ Rods aligned perpendicular to the director: POM photomicrographs of a DMOAP coated (a) nanorod and (b) microrod oriented perpendicular to the director. (c,d) Photomicrographs of the same rods with a $\lambda$-plate (slow-axis) oriented at 45$^{\circ}$. Double-headed arrow at the right bottom indicates director ${\bf {\hat{n}}}$. Director configuration around a micro or nanorod.
 \label{fig:fig3} }
\end{figure}

In a planar cell, the DMOAP coated micro and nanorods are mostly oriented  either parallel or perpendicular to the director ${\bf \hat{n}}$.  Fig.\ref{fig:fig2} (a,b) shows the polarising optical microscope (POM) images of the rods oriented parallel to the director.  The parallel rods create a strong elastic distortion in the surrounding medium and the induced defect by the rods is atypical as compared to the spherical microparticles. In particular, the hyperbolic hedgehog defect ($s=-1$) is not observed near the particle. However, the $\lambda$-plate images indicate that the director is rotated anti-clockwise (yellow region) and clock-wise (blue region), above and below the rods, respectively with respect to the rubbing direction (Fig.\ref{fig:fig2} (c,d)). This means the induced defects are virtual which resides within the rods. Fig.\ref{fig:fig3} (a,b) shows POM images of a perpendicular micro and nanorod. In this case the elastic distortion is minimum and it is hard to see the rods under POM or with $\lambda$-plate. Overall, for both orientations of the rods, the defect structure is unclear from the optical images. 
However, from the mutual elastic interaction it has been shown that the parallel rods have dipolar and perpendicular rods have quadrupolar type director configurations~\textcolor{blue}{\cite{Rasi}}. 
Topological defects of DMOAP coated bigger microrods ($L\simeq8~\upmu$m and $D\simeq1.5~\upmu$m) was studied by Tkalec \textit{et al.}~~\textcolor{blue}{\cite{ut}} and they observed  microrods parallel to director induce point defects and perpendicular microrods induce ring-defects along it's length. 
Aya \textit{et al.} have studied even bigger microrods ($L\simeq60~\upmu$m and $D\simeq10~\upmu$m) dispersed in azodendrimer doped NLC which provides weak homeotropic surface anchoring~\textcolor{blue}{\cite{aya}}. The defect structure of our rods appears to be nontrivial and similar to those reported by Aya \textit{et al}~\textcolor{blue}{\cite{aya}} which implies the surface anchoring of the rods in NLC is weak.

\begin{figure}
\center\includegraphics[scale=0.24]{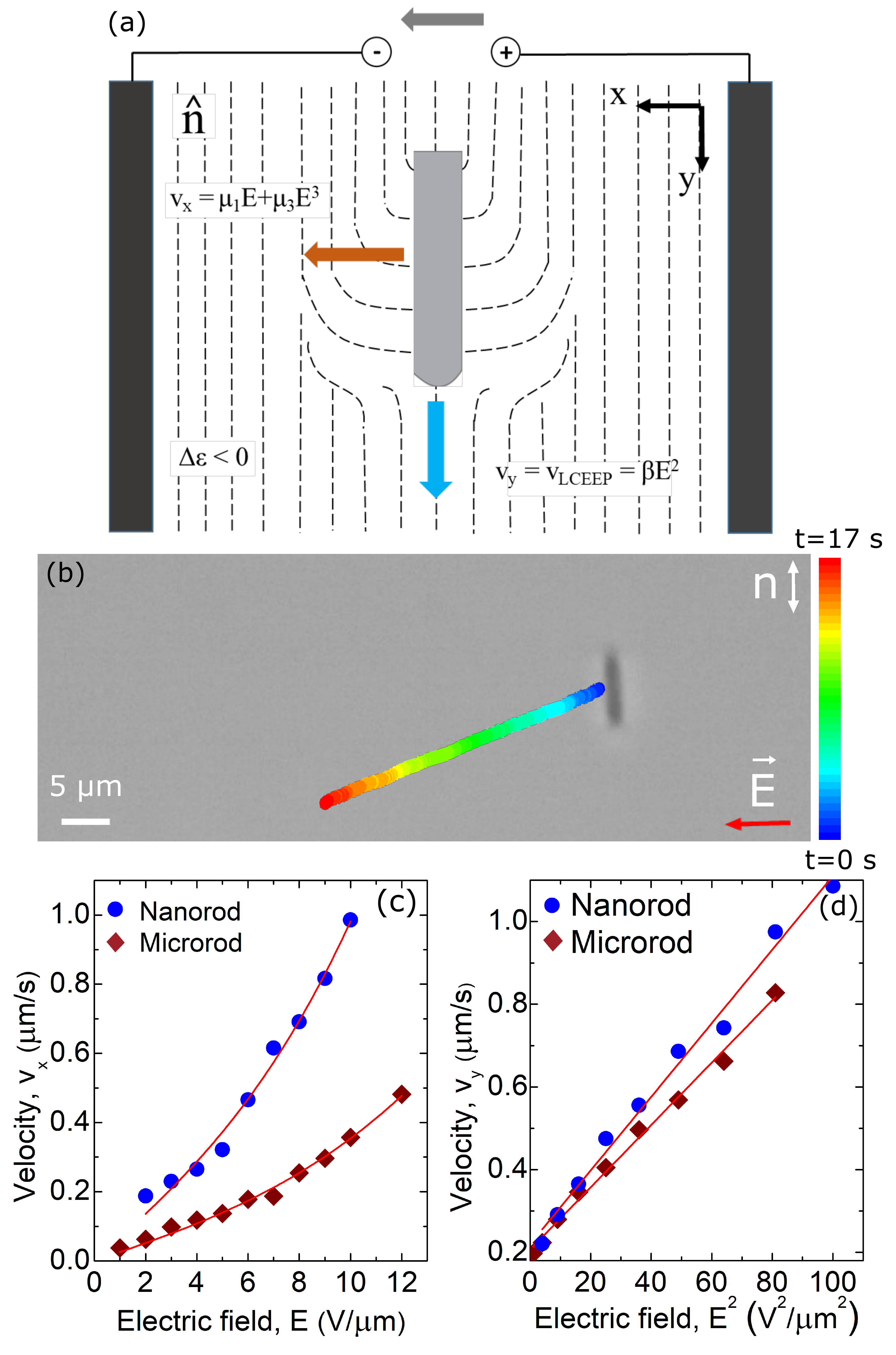}
\caption{(a) Schematic diagram of a planar cell with in-plane stripe electrodes for DC electrophoresis. (b) Colour coded trajectory of a microrod. (c) Variation of $v_x$ and nonlinear least square fits (solid curves) to Eq.(2). (d) Variation of $v_y$  and fits to Eq.(5).  Fit parameters are presented in Table-\ref{table:2}. (see Movie S1 in supplementary)
\label{fig:fig4}}
\end{figure}

  In electric field experiments, first we study the DC electrophoresis of the rods in cells with in-plane stripe electrodes. A schematic diagram of the cell is shown in Fig.\ref{fig:fig4}(a).
  The rods propel at an angle with respect to the electric field (see Fig.\ref{fig:fig4}(b)). It has two velocity components, $v_x$ and $v_y$. The velocity component $v_y$ arises due to the liquid crystal-enabled electrophoresis and we will discuss it later. Since the rods have surface charge the component $v_x$ arises due to the classical electrophoresis and the field dependent velocity can be expressed as ~\textcolor{blue}{\cite{fv}}
 \begin{equation}
v_x=\upmu_{1}(E-E_{0})+\upmu_{3}(E-E_{0})^{3},
\end{equation}
 where $E_{0}$ is the threshold field required for the motion and $\upmu_{1}$ and $\upmu_{3}$ are the mobilities of the linear and cubic terms, respectively.
Figure .\ref{fig:fig4}(c) shows the variation of $v_x$ with field. A good fitting of the data is obtained and the fit parameters are presented in Table-\ref{table:2}. 

  \begin{table*}[!htbp]
\centering
\begin{tabular}{ |p{2cm}|p{3.5 cm}|p{3.5 cm}|p{3 cm}| }
\hline
Particles & \multicolumn{1}{c|}{$\upmu_{1}$ (m$^{2}$V$^{-1}$ s$^{-1}$)} &\multicolumn{1}{c|}{$\upmu_{3}$ (m$^{4}$V$^{-3}$ s$^{-1}$)} &  \multicolumn{1}{c|}{$\beta$ (m\textsuperscript{3}V\textsuperscript{-2}s\textsuperscript{-1})} \\ \hline
Nanorod  &$(6.7\pm 0.6) \times 10^{-11}$ & $(3.5\pm 0.7) \times 10^{-19}$ & $(8.9\pm 0.3)\times^{-15}$ \\
 \hline

Microrod &$(2.6\pm 0.2) \times 10^{-11}$ & $(1.1\pm 0.1) \times 10^{-19}$  & $(7.5\pm 0.2)\times^{-15}$ \\
\hline
\end{tabular}
\caption{Fit parameters $\mu_1$ and $\mu_3$ are obtained from  Eq.(2) and $\beta$ is obtained from Eq.(5) for DC electrophoresis.}
\label{table:2}
\end{table*}
 
 We can approximately estimate the linear mobility $\upmu_{1}$ of the micro and nanorods. It can be written as ~\textcolor{blue}{\cite{av,tb}}
\begin{equation}
\upmu_{1}=\frac{\epsilon_{LC}\epsilon_{0}\zeta}{\eta_{LC}},
\end{equation}
 where $\epsilon_{LC}$, $\eta_{LC}$ are the dielectric constant and viscosity of the liquid crystal and $\zeta_p$ is the surface potential of the particles. Considering  $\epsilon_{LC}=7$~\textcolor{blue}{\cite{dinesh}}, $\eta_{LC}=20$ m Pas~\textcolor{blue}{\cite{merck}} and corresponding Zeta potentials (Table-\ref{table:1}) the estimated linear mobilities of the micro and nano rods are given by $\upmu_{1}^{micro}\simeq4.1\times10^{-11}$m\textsuperscript{2}/V.s and $\upmu_{1}^{nano}\simeq5.0\times10^{-11}$m\textsuperscript{2}/V.s, respectively. These estimates are very close to that are obtained from the experiments.

 Assuming the diffusion coefficients and the number density of the positive and negative ions are same, the mobility $\upmu_{3}$ of the cubic term of the rods can be expressed as ~\textcolor{blue}{\cite{sb,av}}
 \begin{equation}
 \upmu_{3}=r_e^2\frac{e~\epsilon_{LC}\epsilon_0}{K_{B}T~\eta_{LC}} \frac{Du^2}{(1+2Du)^2}
 \end{equation}
 where $Du$ is the Dukhin number expressed as $Du=\sigma_{rod}/(\sigma_{LC}.r)$, where $\sigma_{rod}$ and $\sigma_{LC}$ are the electrical conductivities of the silica rods and the liquid crystal, respectively. $r_e$ is the effective radius of a sphere with the same volume as the rod and given by $r_e=(\frac{3}{4}r^2h)^{1/3}$, $r$ and $h$ being the radius and length. Considering, $\sigma_{rod}=10^{-12}$ S m\textsuperscript{-1}~\textcolor{blue}{\cite{fv}}, and $\sigma_{LC}=3.2\times10^{-10}$ S m\textsuperscript{-1}~\textcolor{blue}{\cite{dinesh}}, calculated nonlinear mobilities of the nano and microrods are: $\upmu_{3}^{nano}\simeq 2.1\times10^{-21}$ m\textsuperscript{4}V\textsuperscript{-3} s\textsuperscript{-1} and $\upmu_{3}^{micro}\simeq1.2\times10^{-20}$ m\textsuperscript{4}V\textsuperscript{-3} s\textsuperscript{-1}, respectively. These calculated values are somewhat smaller than those obtained in the experiments. This could arise due to effective radius of a microsphere considered in the calculations and also the effective conductivity of the LC-colloid system may be different than the pristine NLC.   
 As we mentioned the $y$ component of velocity of the rods arise due to the liquid crystal-enabled electrophoresis (LCEEP) and it can be expressed as~\textcolor{blue}{\cite{oleg1}}
 \begin{equation}
v_y=\beta(E-E_{0})^{2},
\end{equation}	
where  $\beta$ is the mobility of the quadratic term. Figure \ref{fig:fig4}(c) shows the variation of $v_y$ with field. The experimental data is well fitted to this equation and the fit parameters are presented in Table-\ref{table:2}.

 \begin{figure}[!ht]
\center\includegraphics[scale=0.18]{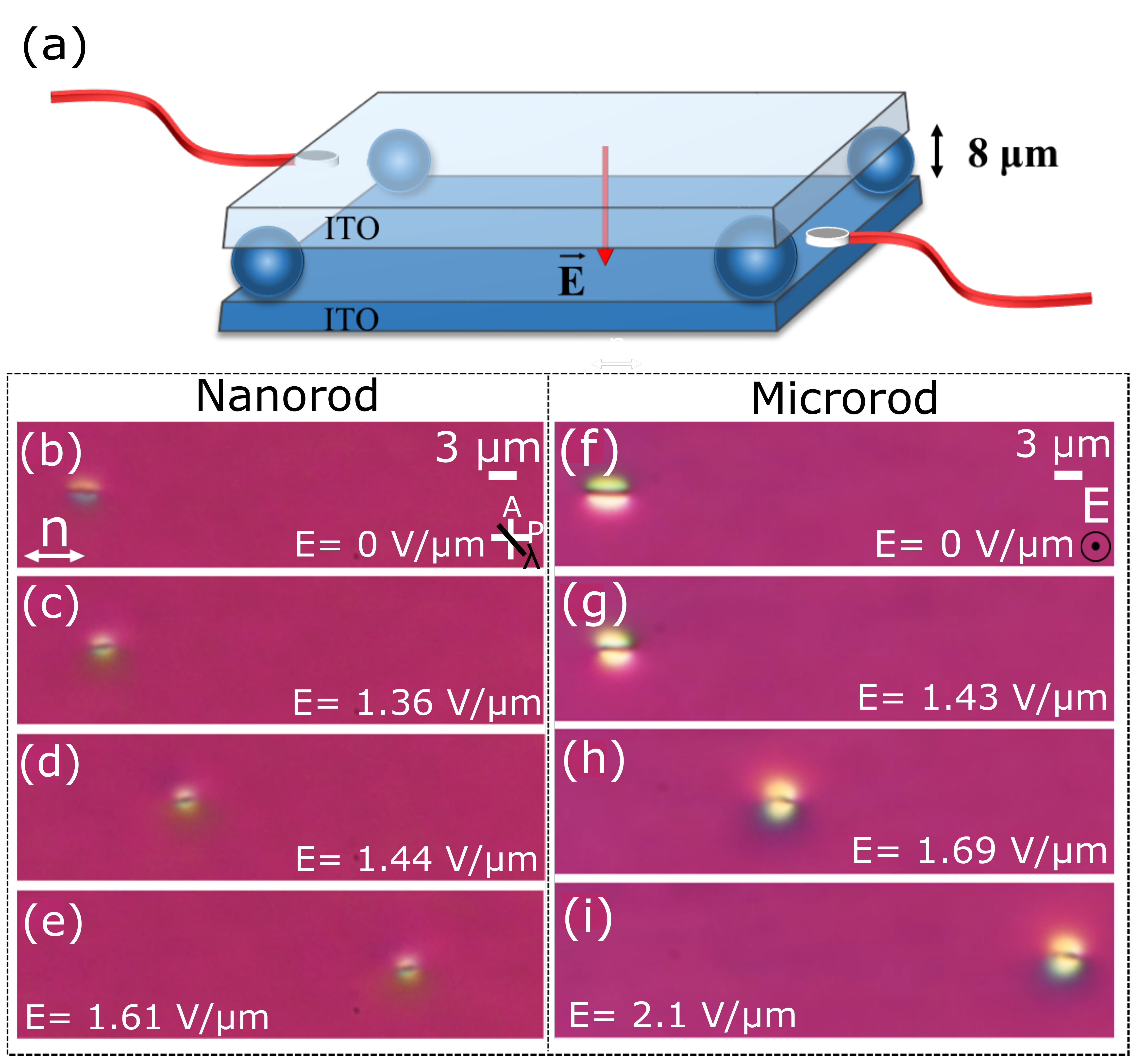}
\caption{(a) Diagram of a Hele-Shaw cell for AC electrophoresis. The direction of the electric field is marked by red arrow which is perpendicular the director ${\bf\hat{n}}$. Effect of AC electric field on a silica (b-e) nano and (f-i) microrod in MLC-6608. Images captured with additional $\lambda$ plate in POM. Frequency=50 Hz. Cell thickness $d=8.5~\upmu$m. (see Movie S2 and Movie S3 in supplementary)
\label{fig:fig5}}
\end{figure}

\begin{figure}[!ht]
\center\includegraphics[scale=0.35]{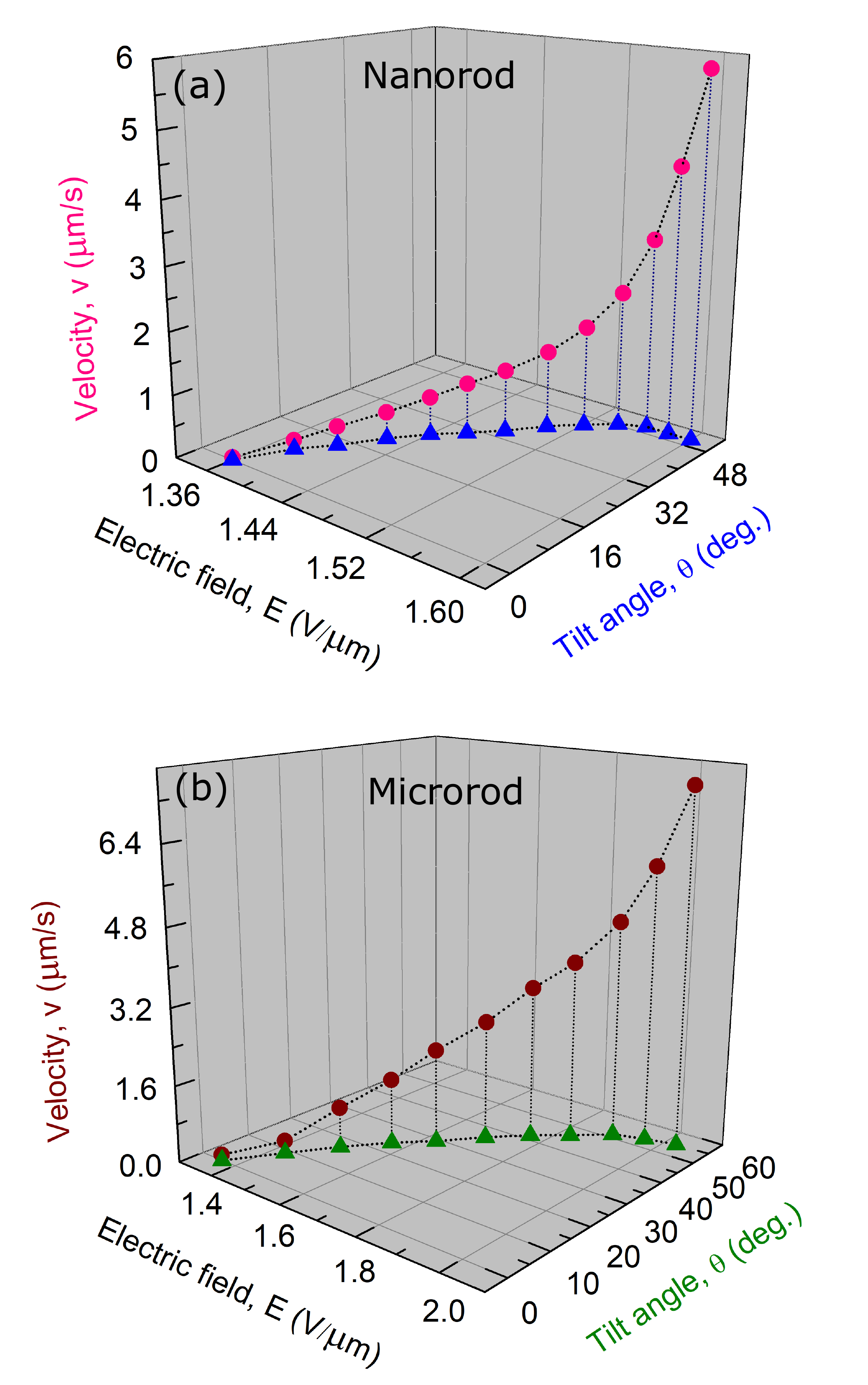}
\caption{ 3D plots of AC electrophoretic velocity (solid circles) with electric field ($E$) and tilt angle $\theta$ (triangles) of a silica (a) nanorod and (b) microrod in a Hele-Shaw cell. Frequency $f=50$ Hz. Cell thickness $d=8.5~\upmu$m.
\label{fig:fig6}}
\end{figure}

In what follows, we study the low frequency (10-120 Hz) AC electrophoresis of the rods in a Hele-Shaw cell as shown schematically in Fig.\ref{fig:fig5}(a). The rods which are aligned perpendicular to the director (see Fig.\ref{fig:fig3}) does not propel under AC field as they have quadrupolar-like director structure~\textcolor{blue}{\cite{Rasi}} hence, we studied AC electrophoresis of rods aligned parallel to the director (see Movie S2 and Movie S3). A sequence of images show the motion of a nano and a microrod with increasing AC electric field. At zero field, the long axes of both the rods are parallel to the rubbing direction (Fig.\ref{fig:fig5} (b) and Fig.\ref{fig:fig5} (f)). When the field is increased, beyond a threshold value ($E_0$=1.36 V/$\upmu$m) the nanorod tilts towards the field direction (perpendicular to the viewing plane) and starts  propelling along the rubbing direction (Fig.\ref{fig:fig5}(c-e)). Similarly, the microrod tilts beyond a slightly higher threshold field ($E_0$=1.43 V/$\upmu$m) and propels along the rubbing direction ((Fig.\ref{fig:fig5}(g-i)). The parallel rods have dipolar-like director structure hence the AC electrophoretic propulsion of such rods is due to the liquid crystal-enabled electrophoresis (LCEEP) as discussed in previously. The frequency dependent velocity of the rods also show a typical response of a dipolar spherical particle (see Supplementary material). We envisage the fore-aft symmetry of electroosmotic flows surrounding the rod is broken due to asymmetric director deformation and the rod propel along the director ($\hat{\bf{n}}$). Here, the applied electric field has two contributions. Firstly, it induces electrophoretic propulsion and secondly, it provides an electrostatic torque that tilts the rods towards the electric field direction.   \\

\begin{figure}[!ht]
\center\includegraphics[scale=0.35]{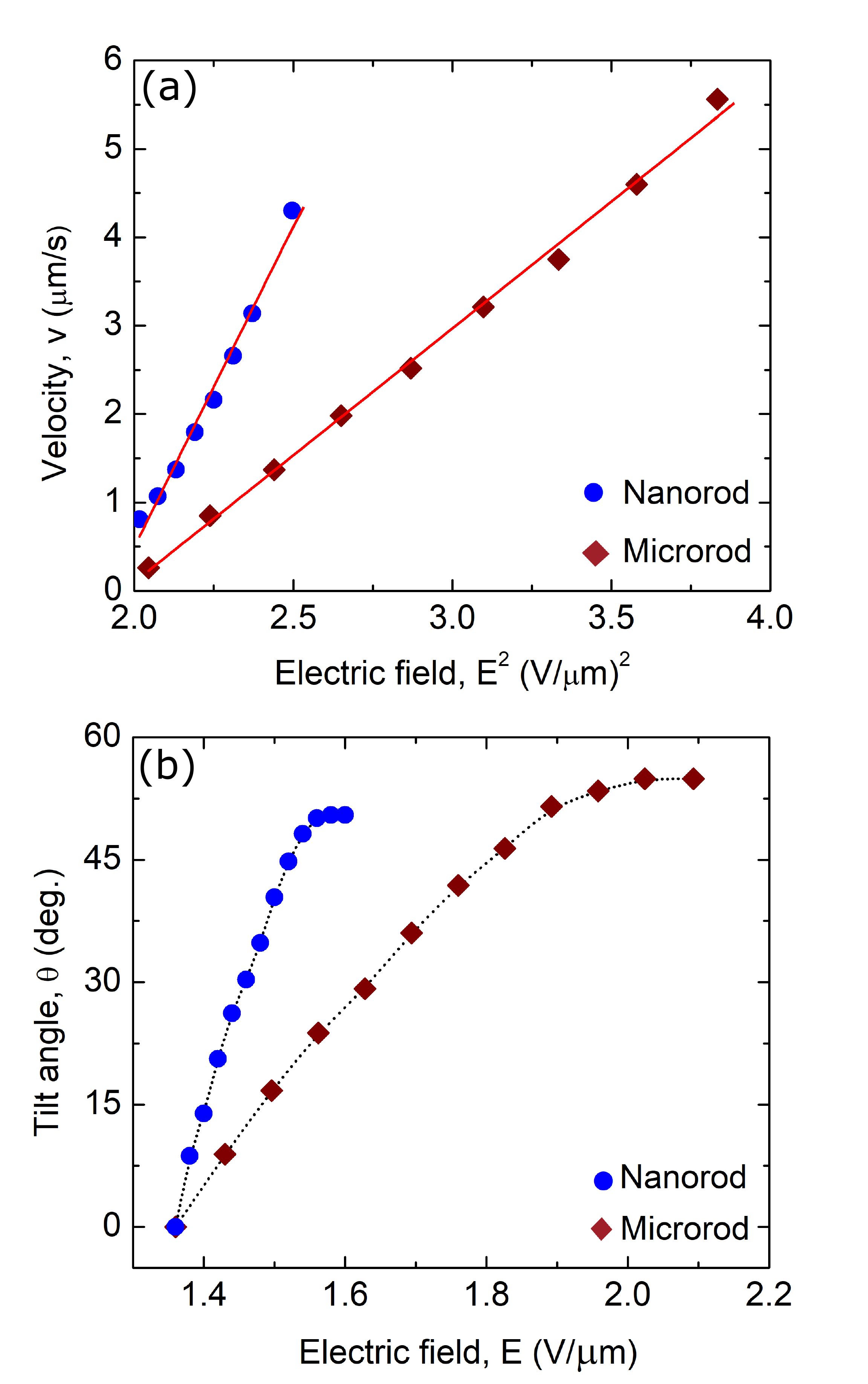}
\caption{(a) Variation of AC electrophoretic velocity $v$ with field of silica nanorods (solid circles) and microrods (solid diamonds). Solid red lines  are nonlinear least square fits to Eq.(5). (b) Variation of tilt angle $\theta$  with  electric field  ($E$) of a nano-rods (solid blue circles) and micro-rods (solid red diamonds). Dotted lines are guide to the eyes. Frequency $f$= 50 Hz. Cell thickness $d=8.5~\upmu$m. 
\label{fig:fig7}}
\end{figure}

We simultaneously measured AC electrophoretic velocity $v$ as well as the tilt angle $\theta$ (angle between the long axis of the rod and the rubbing direction) as a function of applied field ($E$) at a fixed frequency. The tilt angle ($\theta$) of the rods was measured using the equation $\theta=\cos^{-1}(l/L)$, where $l$ and $L$ are the horizontally projected (measured from the movie) and the actual lengths.
Figures \ref{fig:fig6}(a,b) show 3D plots of field dependent  velocity and tilt angle of a nano and a microrod. Figure \ref{fig:fig7}(a) shows that the velocity of both the rods increases quadratically with $E$ and it can be fitted to Eq.(5) with fit parameters $\beta$=(7.1$\pm 0.3$)$\times^{-18}$ m\textsuperscript{3}V\textsuperscript{-2}s\textsuperscript{-1} and (2.9$\pm 0.1$)$\times^{-18}$ m\textsuperscript{3}V\textsuperscript{-2}s\textsuperscript{-1} for the nano and microrods, respectively. These values are closer to that were reported for spherical dipolar particles~\cite{dinesh2}.

Figure \ref{fig:fig7}(b) shows the variation of tilt angle ($\theta$) with applied electric field. The tilt angle of both the rods increase with field and finally saturates around 50$^{\circ}$. 
The tilting of the rods  can be explained based on the competing effect of the field induced electrostatic torque ($\tau_{elec}$) of the rods and elastic torque ($\tau_{elas}$) of the nematic liquid crystal. From the induced dipole moment theory the electrostatic torque exerted on the rod can be written as~\textcolor{blue}{\cite{im,nl,Rasna}}
\begin{equation}
\tau_{el}=\dfrac{1}{2} \alpha_{eff} E^{2} cos\theta sin \theta
\end{equation}\\
where $\theta$ is the angle relative to the director in response to the electric field, $\alpha_{eff}=Re[\alpha^{*}_{long}-\alpha^{*}_{trans}]$ is the real part of the effective polarizability of the rods and $\alpha^{*}_{long}$, $\alpha^{*}_{trans}$ are the complex polarisabilities along the longitudinal axis and transverse axis, respectively. Subsequently the elastic energy exerted on the rod in a NLC increases with the tilt angle and according to  Brochard and de Gennes it can be expressed as ~\textcolor{blue}{\cite{Brochard,le}}

\begin{equation}
F = 2\pi C K \theta^{2}       
\end{equation}
where \textit{C} = $(L/2)$ log(\textit{L}/\textit{D}), \textit{L} and \textit{D} being the length and the diameter of a cylindrical rod.
The elastic torque of the rods can be written as:
$\tau_{elas} =-\frac{\partial F}{\partial \theta}= -4\pi C K \theta$   
At equilibrium the electric torque is balanced with the elastic torque (i.e., $\tau_{elec}$+$\tau_{elas}=0$) and we can write
\begin{equation}
\dfrac{\theta}{cos\theta sin \theta} =  \left( \dfrac{\alpha_{eff}}{8\pi C K} \right)E^{2}
\end{equation}

\begin{figure}[!ht]
\center\includegraphics[scale=0.35]{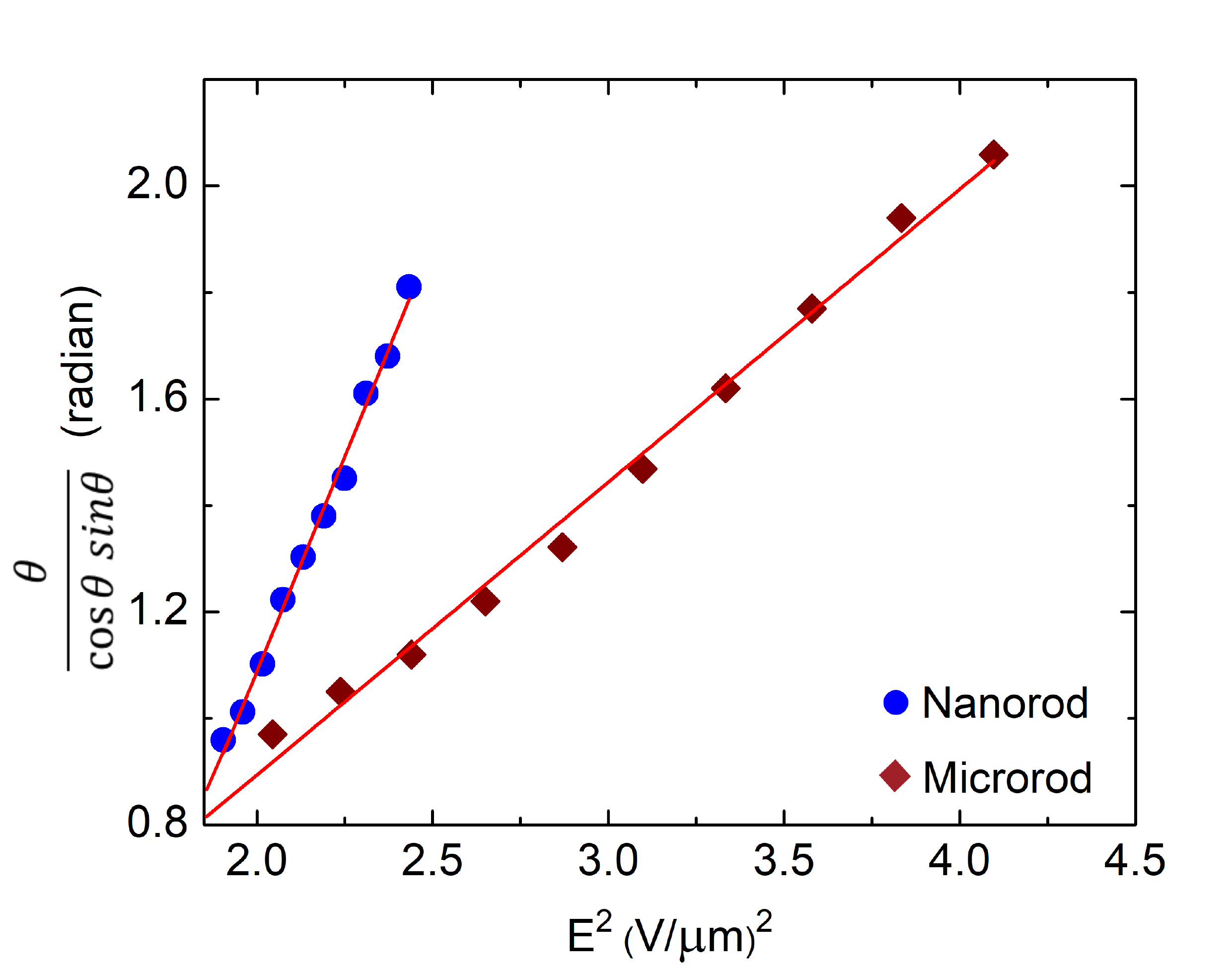}
\caption{Variation of $\theta/cos\theta sin\theta$ with \textit{E}$^{2}$. Red lines are the best fit to Eq.(8). The slopes are 0.55 rad (m/V)\textsuperscript{2} and 1.6 rad (m/V)\textsuperscript{2} for the nano and microrods, respectively. 
\label{fig:fig8}}
\end{figure}

\noindent Figure \ref{fig:fig8} shows the variation $\theta/cos \theta sin \theta$ with $E^{2}$.  The computed quantity on the left hand side of Eq.(8) varies linearly with $E^2$ and a very good fitting is obtained. 
 Considering the average Frank elastic constant \textit{K}=14 $\times$10$^{-12}$ N,  we calculate the effective polarizability, $\alpha^{nano}_{eff}$= (1.2$\pm 0.1$)$\times$10$^{-21}$ C m\textsuperscript{2} V\textsuperscript{-1} and $\alpha^{micro}_{eff}$= (0.52 $\pm 0.1$)$\times$10$^{-21}$ C m\textsuperscript{2} V\textsuperscript{-1} for the nano and microrods, respectively. 
The effective polarizability of the nanorods is almost double than that of the microrods (i.e., $\alpha^{nano}_{eff}/\alpha^{micro}_{eff}\approx 2$).  This is expected as the effective polarizability of electrically polarisable particles increase with the aspect ratio $L/D$ ~\textcolor{blue}{\cite{tt}} and for our nanorods the aspect ratio is almost double compared to that of the microrods. 

We would like to make a comment here. In our system the Reynolds number $Re=\rho v a/\eta\simeq10^{-7}$ and Ericksen number $Er=\eta v a/K\simeq10^{-2}$~\cite{dinesh2}, hence the fluid inertia is negligibly small and the director configuration around the particles remain unaffected by the flow. Also, as in Ref.\cite{oleg2}, we work with zero Peclet number, i.e.,  charge current in our system is diffusion dominated and not affected by the flow. \\\\\\

\section{Conclusion} 
To summarise, we have studied electrophoretic propulsion of two silica rods with different length-to-diameter aspect ratios in a nematic liquid crystal. In contrary to the spherical particles micro or nanorods induce virtual defect i.e., the defects reside within the rods and demonstrate a nontrivial electrophoresis.
  The motion parallel to the electric field is due to the classical electrophoresis and the velocity is described as a sum of linear and cubic terms of the electric field. The linear ($\upmu_{1}$) and nonlinear ($\upmu_{3}$) mobility coefficients measured from the experiments agree well with the calculated values. The motion perpendicular to the field direction is due to the liquid-crystal enabled electrophoresis (LCEEP) and the corresponding velocity is proportional to the square of the electric field.  In  AC electrophoresis the rods experience an electrostatic torque, and an opposing elastic torque, resulting a tilted orientation. The effective polarizability of the nanorods obtained from the torque balance is almost double compared to the microrods as the nanorods have twice the larger aspect ratio than the microrods. Our experiments demonstrate a novel method of measuring linear and nonlinear electrophoretic mobilities and effective polarisability simultaneously. The method is simple and applicable to other shape asymmetric polarisable particles.\\\\

{\bf Acknowledgments:}
AS acknowledges PMRF fellowship, Ministry of Education. 
M.R.M acknowledges CSIR fellowship. SD acknowledges the support from the IoE (UoH/IoE/RC1-20-010), Department of Science and Technology, Govt. of India (DST/SJF/PSA-02/2014-2015). Ravi Kumar Pujala acknowledges SERB Core Research Grant [CRG/2020/006281]. \\\\

{\bf Appendix A.} Supplementary material
\newline
Supplementary data to this article can be found online


\begin{thebibliography}{99}

\bibitem{drop} S. V. Dorp,  U. F. Keyser, N. H. Dekker, C. Dekker, and S. G. Lemay, Origin of the electrophoretic force on DNA in solid-state nanopores, Nat. Phys. {\bf 5}, 347 (2009).


\bibitem{alex} A. Terray, J. Oakey, and D. W. M. Marr, Microfluidic control using colloidal devices, Science {\bf 296}, 1841 (2002).

\bibitem{div2} A. F. Demir{\"o}rs, F. Eichenseher, M. J. Loessner,  and A. R. Studart, Colloidal shuttles for programmable cargo transport, Nat. Commun. {\bf 8}, 1872 (2017).


\bibitem{rc} R. C. Hayward, D. A. Saville, and I. A. Aksay, Electrophoretic assembly of colloidal crystals with optically tunable micropatterns, Nature {\bf 404}, 56 (2000).


\bibitem{cb} C. Bechinger, R. D. Leonardo, H. L{\"o}wen, C. Reichhardt, G. Volpe, and G. Volpe, Active particles in complex and crowded environments, \textit{Rev. Mod. Phys.},  {\bf 88}, 045006 (2016).

\bibitem{sje} S. J. Ebbens, Active colloids: Progress and challenges towards realising autonomous applications, Current Opinion in Colloid \& Interface Science, {\bf 21}, 14 (2016).


\bibitem{rg1}R. Golestanian, T.B. Liverpool, A. Ajdari, Designing phoretic micro- and nano-swimmers, New J. Phys., {\bf 9} 126 (2007).

\bibitem{rg2} J. Howse, R. Jones, A. Ryan, T. Gough, R. Vafabakhsh, R. Golestanian, Self-motile colloidal particles: from directed propulsion to random walk, Phys. Rev. Lett, {\bf 99}, 048102 (2007).


\bibitem{morgan} H. Morgan and N. G. Green, \textit{AC Electrokinetics: Colloids and Nanoparticles} (Research Studies Press Ltd, 2003).

\bibitem{ramos} A. Ramos, \textit{Electrokinetics and Electrohydrodynamics in Microsystems} (Spinger, 2011).


\bibitem{na} N. A. Mishchuk, N. O. Barinova, Theoretical and experimental study of nonlinear electrophoresis, Colloid J. {\bf 73}, 88 (2011).

\bibitem{s} A. S. Dukhin, S. S. Dukhin, Aperiodic capillary electrophoresis method using an alternating current electric field for separation of macromolecules,  Electrophoresis {\bf 26}, 2149 (2005).

\bibitem{t} S. Stotz, Field dependence of the electrophoretic mobility of particles suspended in low-conductivity liquids, J. Colloid Interface Sci. {\bf 65}, 118 (1978).

\bibitem{tm} T. M. Squires and S. R. Quake, Microfluidics: Fluid physics at the nanolitre scale, Rev. Mod. Phys. {\bf 77}, 977 (2005).

\bibitem{baz1} M. Z. Bazant, M. S. Kilic, B. D. Storey, and A. Ajdari, Towards an understanding of induced-charge electrokinetics at large applied voltages in concentrated solutions, Adv. Colloid Interface Sci. {\bf 152}, 48 (2009).

\bibitem{baz2} T. M. Squires and M. Z. Bazant, Induced-charge electro-osmosis, J. Fluid Mech. {\bf 509}, 217 (2004).

\bibitem{baz3} M.Z. Bazant and T. M. Squires, Induced-Charge Electrokinetic Phenomena: Theory and Microfluidic Applications, Phys. Rev. Lett. {\bf 92}, 066101 (2004).

\bibitem{im1} I. Mu\v{s}evi\v{c}, \textit{Liquid crystal colloids}. 2017. (Springer Inter- national Publishing AG). 

\bibitem{is1} I. I. Smalyukh, Liquid crystal colloids, \textit{Annu. Rev. Condens. Matter Phys.},  {\bf 9}, 207 (2018).

\bibitem{oleg1} O. D. Lavrentovich, I. Lazo, and O. P. Pishnyak, Nonlinear electrophoresis of dielectric and metal spheres in a nematic liquid crystal, Nature {\bf 467}, 947 (2010).

\bibitem{oleg} I. Lazo and O. D. Lavrentovich, Liquid-crystal-enabled electrophoresis of spheres in a nematic medium with negative dielectric anisotropy, Phil. Trans. Soc. A {\bf 371}, 2012255 (2013).

\bibitem{od} O. D. Lavrentovich, Active colloids in liquid crystals, Curr. Opin. Colloid Interface Sci. {\bf 21}, 97 (2016).

\bibitem{od1}O. D. Lavrentovich, Transport of particles in liquid crystals, Soft Matter {\bf 10}, 1264 (2014).

\bibitem{oleg2} I. Lazo, C. Peng, J. Xiang, S. V. Shiyanovskii, and O. D. Lavrentovich, Liquid crystal-enabled electro-osmosis through spatial charge separation in distorted regions as a novel mechanism of electrokinetics, Nat. Commun. {\bf 5}, 5033 (2014).

\bibitem{Kuijk} A. Kuijk, A. van Blaaderen and A. Imhof, Synthesis of monodisperse, rodlike silica colloids with tunable aspect ratio, J. Am. Chem. Soc. \textbf{133}, 2346 (2011).

\bibitem{Rasi}M. Rasi, R. K. Pujala and S. Dhara, Colloidal analogues of polymer chains, ribbons and 2D crystals employing orientations and interactions of nano-rods dispersed in a nematic liquid crystal, Sci. Rep., \textbf{9}, 4652 (2019).

\bibitem{Zuhail} K. P. Zuhail, and S. Dhara, Temperature dependence of equilibrium separation and lattice parameters of nematic boojum-colloids,  Appl. Phys. Lett. \textbf{106}, 211901 (2015).

\bibitem{Zuhail1} K. P. Zuhail, and S. Dhara, Effect of temperature and electric field on 2D nematic colloidal crystals stabilised by vortex-like topological defects, Soft Matter \textbf{12}, 6812 (2016).


\bibitem{aya} S. Aya, J. Jougo, F. Araoka, O. Haba and K. Yonetake, Nontrivial topological defects of micro-rods immersed in nematics and their phototuning, Phys. Chem. Chem. Phys. {\bf 24}, 3338 (2022).

\bibitem{ut} U. Tkalec, M. Skarabot and I. Musevic, Interactions of micro-rods in a thin layer of a nematic liquid crystal, Soft Matter {\bf 4}, 2402 (2008).



\bibitem{fv} F. V. Podgornov, A. V. Ryzhkova, W. Haase, Dynamics of nonlinear electrophoretic motion of dielectric microparticles in nematic liquid crystal, J. Mol. Liq. \textbf {267}, 345 (2018).

\bibitem{av}  A. V. Ryzhkova, F. V. Podgornov and W. Haase, Nonlinear electrophoretic motion of dielectric microparticles in nematic liquid crystals, Appl. Phys. Lett. \textbf {96}, 151901 (2010).


\bibitem{tb} T. B. Jones, \textit{Electromechanics of Particles} (Cambridge University Press- 1995, Cambridge).


\bibitem{dinesh} D. K. Sahu, S. Dhara, Measuring electric-field-induced dipole moments of metal-dielectric Janus particles in a nematic liquid crystal, Phys. Rev. Appl., \textbf {14}, 034004 (2020).

\bibitem{merck} Obtained from Merck data sheet on physical properties of MLC-6608 and considered $\eta_{LC}\simeq\eta_2$ (https://www.cmst.be/publi/docdc/doctoraatse27.xml).

\bibitem{sb}  S. Barany, F. Madai and V. Shilov, Study of nonlinear electrophoresis, Appl. Prog. Colloid Polym. Sci., \textbf {128}, 14 (2004).

\bibitem{dinesh2} D. K. Sahu and S. Dhara, Electrophoresis of metal-dielectric Janus particles with dipolar director symmetry in nematic liquid crystals, Soft Matter, \textbf {18}, 1819 (2022).
 

\bibitem{im} I. Minoura and E. Muto, Dielectric Measurement of Individual Microtubules Using the Electroorientation Method, Biophysical Journal  \textbf {90}, 3739 (2006).

\bibitem{nl} P. Kang, X. Serey, Y. F. Chen and D. Erickson. Angular orientation of nanorods using nanophotonic tweezers. Nano Lett., {\bf 12}, 6400 (2012).
\bibitem{Rasna} M. V. Rasna, K. P. Zuhail, U. V. Ramudu, R. Chandrasekar and S. Dhara, Dynamics of electro-orientation of birefringent microsheets in isotropic and nematic liquid crystals, Phys. Rev. E \textbf{94}, 032701 (2016).

\bibitem{Brochard} F. Brochard and P. G. de Gennes, Theory of magnetic suspensions in liquid crystals, J. Physics (Paris) \textbf{31}, 691 (1970).

\bibitem{le} C. Lapointe, A. Hultgren, D. M. Silevitch, E. J. Felton, D. H. Reich, R. L. Leheny, Elastic torque and the levitation of metal wires by a nematic liquid crystal, Science {\bf 303}, 652 (2004).

%\bibitem{mw} M. Washizu, M. Shikida, S. Aizawa and H. Hotani, Orientation and transformation of flagella in electrostatic field, IEEE Transactions and Industry Applications, {\bf 28}, 1194 (1992).


%\bibitem{sim} I. N. Simonov, V. N. Shilov, Theory of low-frequency dielectric-dispersion of a suspension of ideally polarizable spherical particles, Colloid J. USSR  {\bf 39}, 775 (1977).

%\bibitem{ramos1} A. Ramos, H. Morgan, N. G. Green, A. Castellanos, AC electric-field-induced fluid flow in microelectrodes, J . Colloid Interface Sci. {\bf 217}, 420 (1999).

\bibitem{tt}T. Troppenz, A. Kuijk, A. Imhof, A. von Blaaderen, M. Dijkstra and R. van Roij, Nematic ordering of polarizable colloidal rods in an external electric field: theory and experiment, Phys. Chem.Chem. Phys., {\bf 17}, 22423 (2015).





\end {thebibliography}
\end{document}